\documentclass[preprint,showpacs,showkeys,aps,prb]{revtex4}
\usepackage[dvips]{graphicx}

\begin{document}

\title{
Time-Reversible Thermodynamic Irreversibility : \\
 One-Dimensional Heat-Conducting Oscillators \\
and Two-Dimensional Newtonian Shockwaves \\
}

\author{
William Graham Hoover and Carol Griswold Hoover \\
Ruby Valley Research Institute                  \\
Highway Contract 60, Box 601                    \\
Ruby Valley, Nevada 89833 ;                     \\
}
\date{\today}

\vspace{0.1cm}

\begin{abstract}

We analyze the time-reversible mechanics of two irreversible simulation types.
The first is a dissipative one-dimensional heat-conducting oscillator exposed to
a temperature gradient in a three-dimensional phase space with coordinate $q$,
momentum $p$, and thermostat control variable $\zeta$. The second type simulates
a conservative two-dimensional $N$-body fluid with $4N$ phase variables $\{q,p\}$
undergoing shock compression. Despite the time-reversibility of each of the three
oscillator equations and all of the $4N$ manybody motion equations both types of
simulation are irreversible, obeying the Second Law of Thermodynamics. But for
different reasons. The irreversible oscillator seeks out an attractive dissipative
limit cycle. The likewise irreversible, but thoroughly conservative, Newtonian
shockwave eventually generates a reversible near-equilibrium pair of rarefaction
fans. Both problem types illustrate interesting features of Lyapunov instability.
This instability results in the exponential growth of small perturbations, $\propto 
e^{\lambda t}$ where $\lambda$ is a ``Lyapunov exponent''.
\end{abstract}

\maketitle

\section{Reversibility of Dissipative and Conservative Mechanics}

Classical mechanics is {\it time-reversible} in the sense that a movie
of the motion, run backwards, obeys exactly the same motion equations
as does the original forward version. Classical mechanics is an excellent
model for conservative systems free of the real-life dissipative effects
of friction, viscosity, and heat conduction. In order to model
dissipative phenomena on an atomistic scale {\it nonequilibrium}
molecular dynamics includes {\it control variables} in the equations
of motion. These variables use feedback to impose local values of the
temperature and pressure which drive nonequilibrium flows. Thousands
of implementations of this approach have been stimulated by Shuichi
Nos\'e's pioneering 1984 work\cite{b1,b2}. We will explore the time
reversibility of an application of his work here.

Nos\'e's 1984 papers generalize Hamiltonian mechanics with a frictional
variable $\zeta$ controlling the kinetic temperature
$T$ of one or more particular degrees of freedom :
$$
\zeta \longrightarrow kT = \langle \ p^2/m = mv^2 \ \rangle \ .
$$
Here $k$ is Boltzmann's constant, $m$ the mass of a particle, and $mv=p$
the momentum of a controlled degree of freedom. For simplicity in what
follows we set both $k$ and $m$ equal to unity. Hoover applied Nos\'e's idea
to the simplest special case, a one-dimensional harmonic oscillator, in
1985\cite{b3}, later extending that work in 1986 with Posch and
Vesely\cite{b4}. Numerical solutions of the thermostatted oscillator's
equations of motion (with $k$ and $m$ unity now) ,
$$
\{ \ \dot q = p \ ; \ \dot p = -q - \zeta p \ ; \ \dot \zeta = p^2 - T \ \},
$$
can be reversed in either of two ways, by [1] changing the signs of the time
and timestep, $t$ and $dt$, or [2] by changing the signs of $p$ and the control
variable $\zeta$. Both ways simply reverse the time dependence of the
coordinate: $q(+t) \rightarrow q(-t)$.

\section{One-Dimensional Heat-Conducting Oscillator}

\begin{figure}
\includegraphics[width=2.in,angle=-90.]{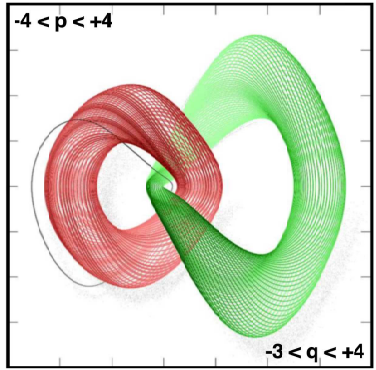}
\caption{
Two conservative tori are interlinked with a dissipative limit cycle\cite{b6}. All three
stationary solutions of the oscillator equations for $T(q) = 1 + 0.42\tanh(q)$
can be generated with initial values $(q,p,\zeta)$  = (-2,-2,0), (-2.3,0,0) and
$(+3.5,0,0)$. If instead $T(q) = 1 + \tanh(q)$ all initial conditions lead to
the limit cycle illustrated in Figures 2-4.
}
\end{figure}

In 1997 Posch and Hoover generalized the oscillator problem, specifying a
coordinate-dependent temperature $T(q) = 1 + \epsilon \tanh(q)$. This temperature
profile has a maximum temperature gradient, $(dT/dq) =\epsilon$ at $q=0$. Particular
choices of $\epsilon$ generated a variety of $(q,p,\zeta)$ ``strange attractors''
[fractal distributions in $(q,p,\zeta)$ space]. Figure 5, a cross-section through a
fractal attractor, gives an impression of the complicated structures resulting from
relatively simple ordinary differential equations. The fractional dimensionalities
of these attractor distributions were all between 2 and 3. Other initial conditions
or choices of $T(q)$ resulted in one-dimensional limit cycles rather than fractals\cite{b5}.
More recently, in 2014, Sprott and the Hoovers found initial conditions, $(q,0,0)$, with
$T(q) = 1 + 0.42\tanh(q)$, which generate two distinct families of conservative
tori (with the initial values $q = -2.3$ and 3.5)\cite{b6}, The tori are interlinked and
coexisting stably with a one-dimensional dissipative limit cycle. The cycle can be 
generated easily with initial values $(q,p,\zeta) =(-2,-2,0)$ and a fourth-order
Runge-Kutta timestep $dt = 0.01$. Figure 2 in Reference 6 shows the three
interlinked phase-space structures. We reproduce it here as Figure 1.

\begin{figure}
\includegraphics[width=2in,angle=-90.]{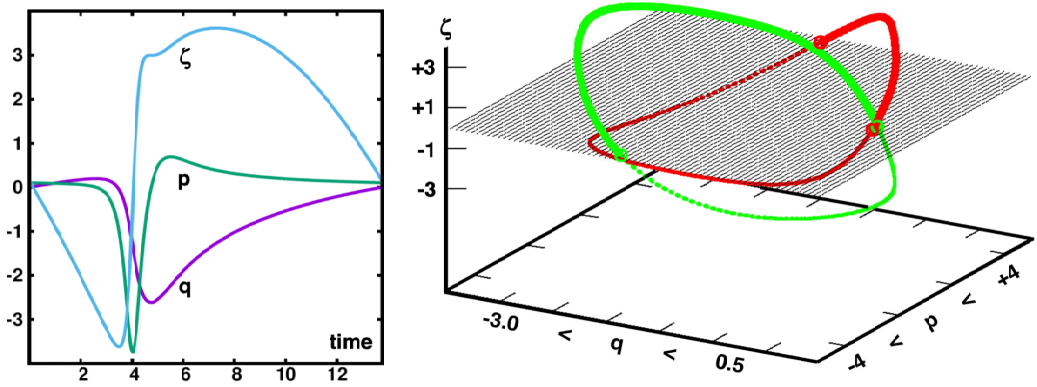}
\caption{
Stable and unstable limit cycles for a heat-conducting one-dimensional
harmonic oscillator. At the left the time dependence of coordinate $q$,
momentum $p$, and friction coefficient $\zeta$, purple, green, and blue
respectively, for the conducting oscillator with period $\tau=13.7494$.
At the right are three-dimensional plots of the attractive orbit (green),
with initial condition $\{ q,p,\zeta \} = \{0, 0.1050726, 0.1455481\}$
and the corresponding {\it unstable} repulsive orbit (red), with $p$ and
$\zeta$ changed in sign. The mean value of the (hot to cold) heat current
is $\langle \ (p^3/2) \ \rangle = -0.74383$. The red and green circles
indicate the four crossings of the orbits with the $\zeta = 0$ grey plane.
}
\end{figure}

In this rich collection of one-dimensional limit cycles, two-dimensional tori,
and fractional-dimensional strange attractors the simplest special case is
arguably $T(q) = 1 + \tanh(q)$. The coordinate-dependent temperature varies from
0 to 2 as $q$ varies from $-\infty$ to $+\infty$. We believe that the basin of
attraction for this case is the entire three-dimensional phase space. To support
this idea we chose a square $200 \times 200$ grid of $(q,p,\zeta)$ points in the
$\zeta=0$ plane with $q$ and $p$ ranging from $-4.975$ to $+4.975$ in steps of
0.05. For each of these 40,000 initial conditions we generated an orbit of length
100$\tau$, and plotted the $(q,p)$ values whenever $\zeta$ changed sign. Every
one of these long orbits ended up crossing at the two penetration points plotted
in green in Figure 2. For each orbit we used a timestep of $\tau/10000$ where
the cycle period is $\tau = 13.7494$. Let us explore that solution in more detail,
based on fourth-order Runge-Kutta numerical simulations.

\begin{figure}
\includegraphics[width=3.in,angle=-90.]{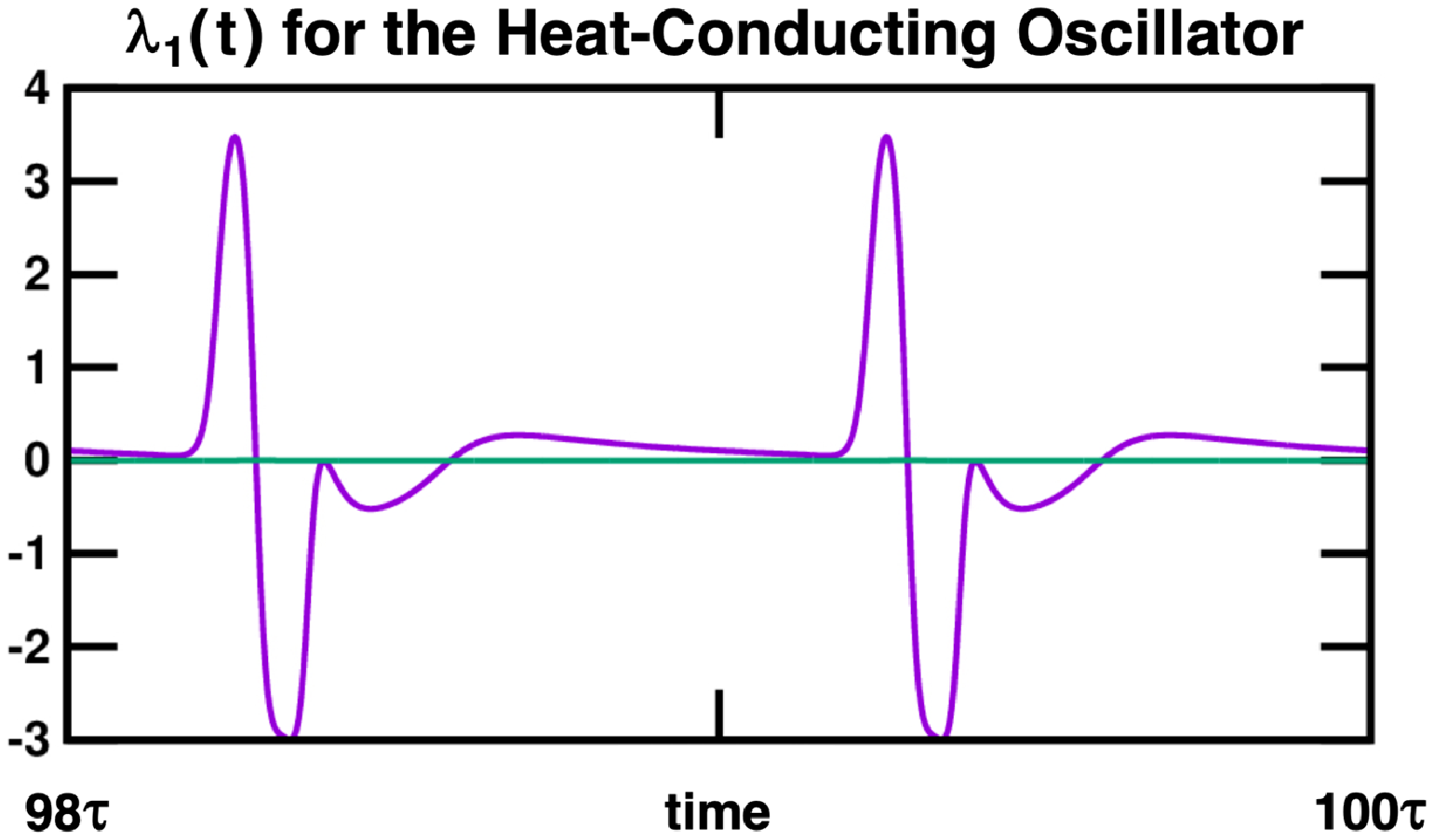}
\caption{
Time dependence of the local Lyapunov exponent for the oscillator during
its 99th and 100th period of oscillation.  The local value is purple.
The green time average of $\lambda_1(t)$ vanishes, corresponding to the lack of averaged
strain along the trajectory.
}
\end{figure}

From the phase-space analog of Liouville's continuity equation the mean value
of the friction coefficient $\zeta$ necessarily corresponds to the time-averaged
loss rate of phase volume, $\otimes = dqdpd\zeta$: 
$$
\langle \ (\dot \otimes/\otimes) \ \rangle = \langle \ (\partial \dot q/\partial q) +
(\partial \dot p/\partial p) + (\partial \dot \zeta/\partial \zeta) \ \rangle =
\langle \ 0-\zeta+0 \ \rangle = -1.325 \ .
$$
In a single period the comoving phase volume decreases by a factor of $e^{\langle \
\zeta \ \rangle \tau} = e^{1.325\times 13.7494} \simeq 10^8$. The maximum temperature
gradient, $(dT/dq)$, is unity, at $q=0$. The mean heat current, averaged over time,
$\langle (p^3/2) \rangle$, is $-0.74383$, and the net transport of kinetic energy
$(p^2/2)$ is from right to left, consistent with thermodynamics' Second Law.

The Lyapunov exponents, three of them in a three-dimensional phase space, measure the
comoving expansion rates of the phase volume $\otimes$ : 
$$
(\dot \otimes/\otimes)= \lambda_1(t) + \lambda_2(t) + \lambda_3(t) \ .
$$
The one-dimensional limit cycle's largest Lyapunov exponent has an average value of
zero, as shown in Figure 3. The vanishing mean value of $\lambda_1$ corresponds to the
averaged lack of relative motion of two adjacent trajectory points along the attractive
one-dimensional trajectory. $\lambda_2$ and $\lambda_3$ have negative averages,
$-0.110$ and $-1.215$, describing the net rates of convergence of nearby 
trajectories in the two directions perpendicular to the limit cycle.

For the special case $T(q) = 1 + \tanh(q)$ the longtime solution of the motion
equations forward in time is the unique attractive periodic orbit shown (green)
in the right panel of Figure 2. The period $\tau$ is 13.7494 with $-2.616 < q <
 +0.198$. The reversed orbit, with the same range of $q$ visited in the opposite
time direction, is half the attractor/repellor pair. The repellor is only
observable briefly due to its inherent Lyapunov instability, proportional to
$e^{\lambda_1(t)\times t} \simeq e^{+1.215t}$. We expect to see the exponential
growth of an original one-step roundoff error grow to observability in just
a few oscillator periods. We examine that next.

\begin{figure}
\includegraphics[width=3.2 in,angle=-90.]{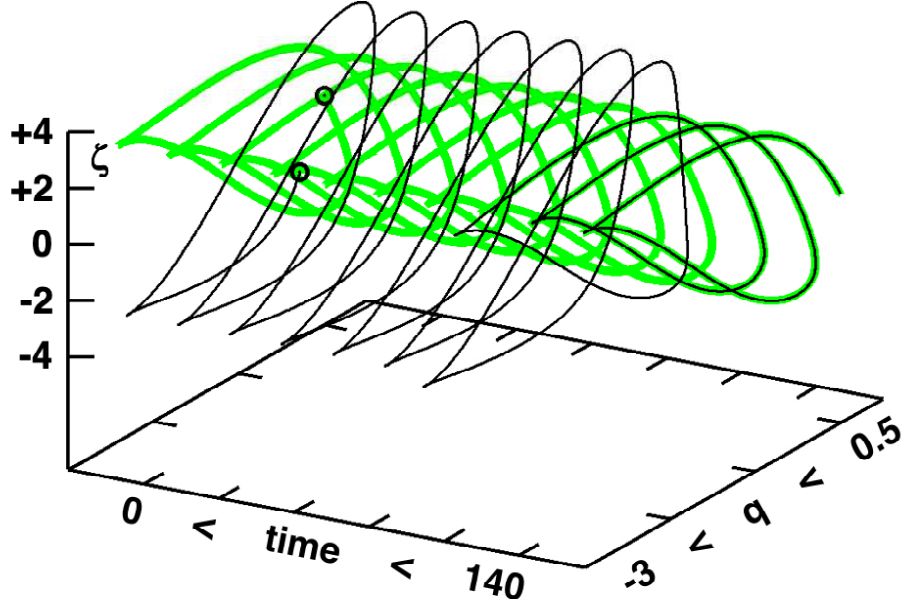}
\caption{
Evolution of ten $(q,\zeta)$ orbits beginning at the initial upper black circle
with the conditions of Figure 2. At time $t = 137.494$ the sign of the timestep
is reversed. The reversed trajectory (black) traces back accurately for nearly
three reversed orbits before making a rapid switch to the stabilized repellor,
following it to the lower black circle at $t=0$.
}
\end{figure}

\section{Stability and Instability of Periodic Oscillator Orbits}

Figure 4 illustrates the stability of the attractive limit-cycle orbit,
$\{ \lambda_i\} = \{ 0,-0.110,-1.215 \}$ and the instability of its time reversal,
$\{ \lambda_i\} = \{ +1.215,+0.110,0 \}$ with a two-stage simulation. First we follow
ten circuits of the attractor in green, using a million quadruple-precision
fourth-order Runge-Kutta timesteps of $dt = 10^{-5}\tau = 0.000137494$,
implying a local integration error of order $dt^5/120 \simeq e^{-49}$ at each
step. We then reverse time, $+dt \rightarrow -dt$, converting the stable attractor
to the unstable reversed repellor. Unlike the stable attractor the repellor is
unstable, with a positive Lyapunov exponent $\lambda_1 = 1.215$. This suggests
that the reversal should become visibly inaccurate at a time of order $49.3/1.215
\simeq$ three or four periods. This agrees well with the reversed black trajectory
of Figure 4, which follows the reversed $(q,p,\zeta)$ orbit (stabilized by the
negative $dt$) for between six and seven periods, ending up at the lower black circle :
$$
\{ \ q = 0, \  p = -0.1050726, \ \zeta =  - 0.1455481 \ \} \ .
$$

The largest Lyapunov exponent is relatively easy to measure. Follow two neighboring
trajectories, the ``reference'' and the ``satellite'', adjusting the satellite location
by rescaling its separation from the reference, $\delta_{t+dt} \rightarrow \delta_{\rm o}$, 
at the conclusion of each time step. This rescaling precisely counters the exponential
growth which would occur in the absence of rescaling. The local Lyapunov exponent follows
from the rescaling operation :
$$
\lambda_1 dt = -\ln(\delta_{\rm after}/\delta_{\rm o}) \rightarrow \lambda_1(t) =
(-1/dt)\ln (\delta_{t+dt}/\delta_{\rm o}) .
$$
A convenient choice for $\delta_{\rm o}$ is $0.00001$. Figure 3 shows the time
variation of the largest Lyapunov exponent $\lambda_1$, which lies in the range
$-3.00 < \lambda_1(t) < +3.48$ with a mean value of zero, corresponding to the
(lack of) growth rate of perturbations parallel to the trajectory.

\section{Time Reversibility and Loschmidt's Paradox}

A classic physics puzzle addresses the surprising coexistence of macroscopic
irreversibility with microscopic time reversibility. In 1876 Loschmidt
pointed out that any solution of the equations of motion which is time
reversible and demonstrates the production of entropy can be made to
violate the Second Law of Thermodynamics by analyzing the reversed
motion. Simply stated, time-reversible mechanics necessarily violates
the Second Law in one of the two time directions. We have already seen
that time-reversible Nos\'e-Hoover mechanics, with control of the kinetic
temperature, obeys the Second Law. This is not only possible, but inevitable,
for computational models of the heat-conducting oscillator. With $dt$ positive,
attractive distributions such as the oscillator limit cycle are inevitably
observed.  Repellors are not, due to their vanishing probabilities. 

The exploration of a simple one-body time-reversible model\cite{b7}, and its
relation to entropy production in a  many-body system with heat flow\cite{b8},
clarified the reversibility paradox in 1987. The ``Galton Staircase''
pictures a reversibly thermostatted particle in a downhill steady state,
driven by a periodic sinusoidal potential superimposed on a constant downhill
field. As the particle falls more than it climbs, the model generates a fractal
(fractional dimensional) phase-space distribution in its three-dimensional
phase space $(q,p,\zeta)$. The resulting zero-volume attractive fractal, when
reversed, corresponds to the extreme rarity of states violating the Second Law of
Thermodynamics by converting heat to work. The mirror-image repulsive fractal,
corresponding to an upward moving particle violating the Law has, like the
attractor, zero volume, but is repulsive and of zero probability. The ``attractor''
{\it is} attractive, with probability one. This difference in behavior occurs because
Lyapunov instability is {\it not} time-reversible.

In the Galton staircase a particle travelling uphill, as described by the repellor
states, violates the Law by converting kinetic energy to potential. Repulsion,
coupled with zero volume, makes these fractal repellor states unobservable.
The conducting oscillator of Figures 2-4 offers a simple analog for a particle
transporting energy from hot to cold rather than transporting mass through motion
driven by a gravitational field. Both systems resolve Loschmidt's Paradox
by introducing time-reversible variables controlling temperature. It
is the extreme unobservable rarity of repellor states, the fractal set for uphill
motion in the Galton Staircase, and here the repulsive one-dimensional repellor
limit cycle in three-dimensional space, that forces motions to obey the Second Law.
Staircase simulations reveal the exponential growth of the separation from the fractal
repellor and an irresistible attraction to the repellor's mirror-image attractor.  Likewise the
conducting oscillator with $T = 1 + \tanh(q)$ follows the attractive limit cycle
of the Figures, rather than the cycle's  mirror image repulsive twin, which repels rather than
attracts.

Holian, Hoover, and Posch\cite{b8} stressed that similar irreversible behavior occurs
in the reversible simulations of thermostatted nonequilibrium manybody systems. They
described a heat-conducting system in contact with two reservoirs, one hot and the
other cold. Such a system loses phase volume when it satisfies Fourier's Law,
transmitting heat from the hot reservoir (with an entropy production $Q/T_H)$ to the
cold (with an entropy loss $Q/T_C)$ which necessarily exceeds the gain. The result is
a phase-volume loss exponential in the time.  Just as in the Galton Staircase mass-flow
problem heat flow {\it from hot to cold} results in fractal phase-space structures. Both 
fractal types have zero volume, with zero probability of observing the repellor and with
inevitable longtime probability one for the attractor.

\section{An Illustrative Fractal for the Conducting Oscillator}

To help visualize the attractors and repellors that characterize nonequilibrium
systems we consider here a fractal resulting from dissipation controlled by a {\it pair}
of control variables. $\xi$ controls the fourth moment of the velocity distribution
while $\zeta$ controls the second.
Because fractal distributions are difficult to visualize in their entirety they
are typically described by projections or cross sections. The four-dimensional
conducting oscillator introduced by Posch and Hoover\cite{b5} provides a variety
of thought-provoking fractal structures. As an example, for the same temperature
profile considered here, $T = 1 + \tanh(q)$, see Figure 5. The additional
phase-space dimension results from using two control variables rather than one.
The doubly-thermostatted oscillator requires the solution of four ordinary
differential equations:
$$
\{ \ \dot q = p \ ; \ \dot p = -q - \xi p^3 - \zeta p \ ; \ \dot \xi = p^4 -3p^2T \
; \ \dot \zeta = p^2 - T \ \} \ .
$$ 
These motion equations are fully ergodic for the special case in which $T=1$.  That
is, all possible values of the four variables occur and with the known distribution:
$$
4\pi^2{\rm prob}(q,p,\xi,\zeta) = e^{-(q^2+p^2+\xi^2+\zeta^2)/2} \ .
$$
\begin{figure}
\includegraphics[width=2.6in,angle=-90.]{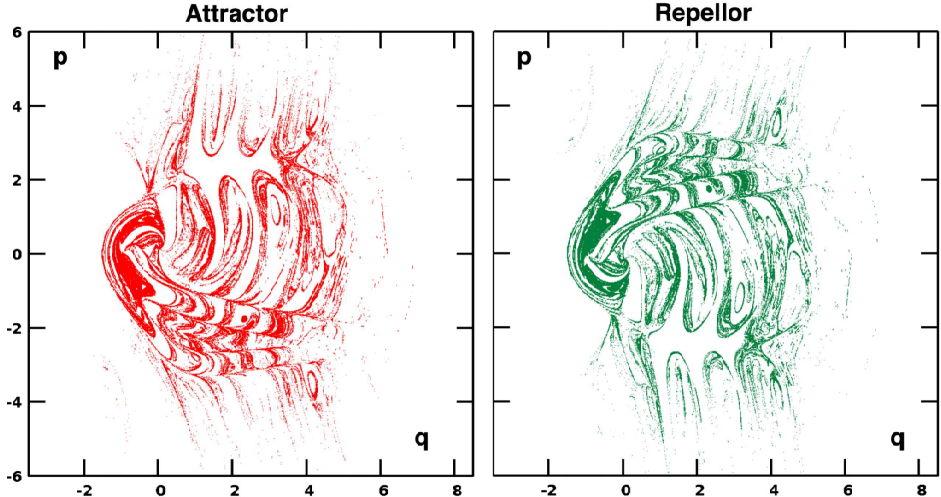}
\caption{
This double cross section is generated by plotting $(q,p)$ trajectory points 
whenever the two thermostat values are near zero. The Figure includes ten million points
satisfying the condition $\xi^2 + \zeta^2 < 0.0001$. Fourth-order Runge-Kutta integration
with $dt = 0.001$ was used.
}
\end{figure}

\begin{figure}
\includegraphics[width=3.3in,angle=-90.]{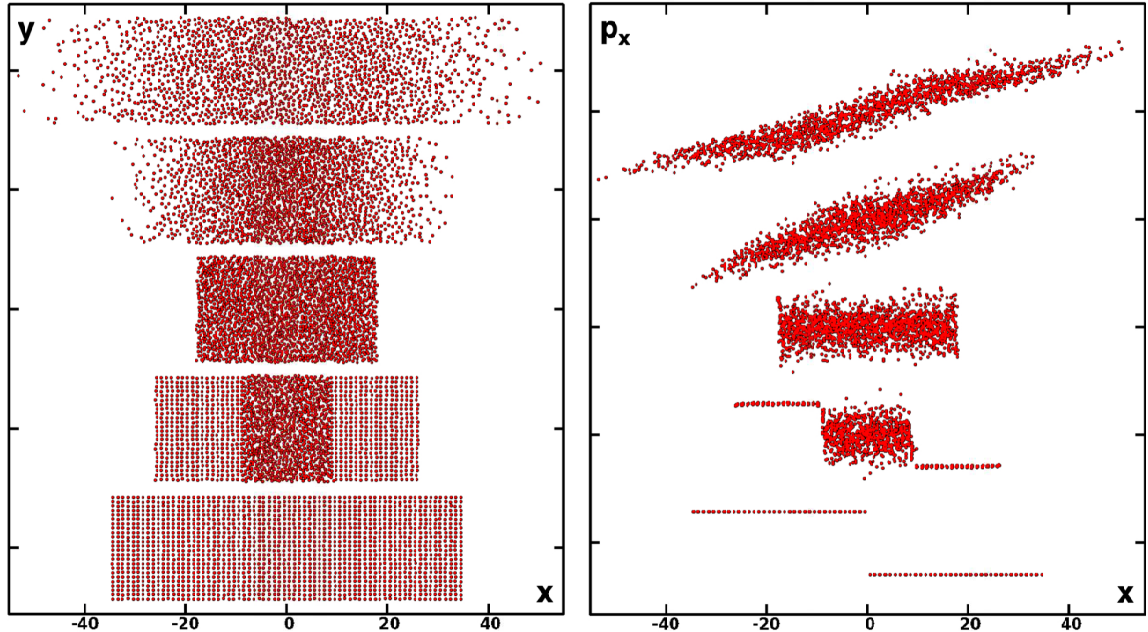}
\caption{
At the left we see five particle plots, at times 0, 10, 20, 30, and 40 from bottom to top,
during the twofold shock compression $(0 < t < 20)$ and subsequent generation of rarefaction waves
$(t > 20)$ simulated by the headon collision of $2\times840= 1680$ soft disks. The initial
square-lattice coordinates (at bottom left) and horizontal velocities $u_p = u_s/2 = \pm 0.875$
(at bottom right) are selected so as to generate twofold shock compression, doubling the density
to generate a hot fluid, reaching it with the coordinates and velocities in the middle view of
Figure 6 and the bottom view of Figure 7. Initially the lattice symmetry is broken with random displacements of $x\
$ and $y$ in the range from $-0.05$ to $+0.05$. The top two plots show
the growing twin rarefaction fans launched from the edges of the hot dense fluid. At the right
the evolving horizontal velocity components are shown at the same five equally-spaced
times, from 0 to 40. The rarefaction fans at times 30 and 40 are expanding at about the
speed of sound. The smooth short-ranged repulsive pair potential governing the Newtonian dynamics is
$\phi(r<1)=(10/\pi)(1-r)^3$. The fourth-order Runge-Kutta integrator with $dt = 0.01$ was used
for Figures 6 and 7.
}
\end{figure}

\section{Two-Dimensional Strong Shockwaves Do Not Reverse}

We have seen that Nos\'e-Hoover temperature control provides a probabilistic mechanism
for irreversibility, the formation of attractor-repellor pairs in phase space which
stabilize the attractor and destabilize the repellor, both through Lyapunov instability.
We have recently detected a related mechanism, but free of control variables, demonstrating the
irreversibility of purely Newtonian mechanical systems, illustrated here with an atomistic
model generating strong shockwaves. Shockwaves are localized regions, usually just a few
atomic diameters in width, within which density, pressure, energy, and temperature all
undergo substantial increases. We consider a model here where the temperature changes by a
factor of 100 and the density doubles. Shockwaves are relatively easily treated computationally
because they are bounded by equilibrium thermodynamic states. To ease the computational burden
we consider the shockwave compression of two-dimensional soft-disk particles in two space
dimensions. The sudden compression occurs in the $x$ direction. The purely-repulsive pair
potential is chosen for simplicity, $\phi(r<1) = (10/\pi)(1-r)^3$. The initial near-zero-pressure
state is a nearly perfect square lattice, with lattice spacing of unity. The small initial
displacements in the range $\pm 0.05$ correspond to a temperature of order 0.001. 

Consider the head-on collision of two $N$-body mirror-image zero-pressure zero-energy
blocks of material with opposite velocities\cite{b9,b10,b11}, shown in Figure 6. Here
$N = 35\times 24 = 840$. With periodic boundaries in $y$ the two colliding $N$-body
blocks steadily convert their kinetic energy to heat. At any stage in the simulation a
reversed solution will show, briefly, antithermodynamic behavior, converting some of the
internal energy of the stagnating blocks back into the original directed kinetic energy,
$(p_x^2/2)=(0.875^2/2)$ per particle.

Berni Alder and Marvin Ross emphasized the highly irreversible nature of 
shockwaves\cite{b12} as follows :
``the most irreversible way to go from one thermodynamic state to another''.
A simple example of this transformation is illustrated in Figure 6, where $70 \times
24 = 1680$ particles undergo twofold compression and then expand to form a symmetric
pair of  rarefaction fans. The initial condition for this example is a neighboring mirror-image
pair of colliding square-lattice blocks, both at the stress-free density of unity. The left
half travels rightward and the right half leftward. Periodic boundaries are imposed in
the $y$ direction, at the top and bottom of the two colliding 840-body blocks. The dynamics is
purely Newtonian. The difference between the steady shockwaves forward in time and
the unsteady rarefaction (rather than shocks) waves in the reversed time direction of
Figure 7 shows that the shockwaves are irreversible. In fact the irreversible
Navier-Stokes equations of motion predict that a reversed shockwave will immediately
widen and slow, transforming into an unsteady rarefaction fan\cite{b9,b10}.

Figure 7 was constructed by reversing the velocities of all particles in Figure 6 at
the time 20, the time of maximum twofold compression. Notice that the snapshot
second from the bottom of Figure 7, where the flow has been reversed so that the
configuration is only halfway to maximum compression, resembles closely that second
from the bottom in Figure 6, where the flow is forward, and halfway to the time of
maximum compression. This apparent reversibility
suggests that the initial single-step Runge-Kutta integration error,
$$
dt = 0.01 \rightarrow 0.01^5/5! = e^{-27.8} \ {\rm or} \
dt = 0.005 \rightarrow 200^{-5}/5! = e^{-31.3} \ ,
$$
expands exponentially to become of order unity at $t = 27.8/\lambda$ or $31.3/\lambda$
in these two typical cases.

Attempting to confirm and elaborate this estimate we constructed reference and satellite trajectories,
for a range of timesteps from $n = 1$ through $n = 7$ :
$$
dt = 0.1/2^n \ ; \ n = 1 \rightarrow dt = 0.05 \ ... \ n=7 \rightarrow dt = 0.00078125 \ ,
$$
rescaling their reference-to-satellite separation to 0.00001 at the end of each timestep. Typical resulting
values of the local-in-time Lyapunov exponent $\lambda_1(t) \simeq 1.8$ are shown in Figures 8 and 9. The
complete set of 6720 two-block motion equations is included in the Lyapunov calculation. The signs of all
the velocity components are changed at time 20 in Figure 8 and at time 10 in Figure 9, with the particle
coordinates pictured as the latter figure's inset.

Fluctuations of the local exponent $\lambda_1(t)$ can be substantially reduced by smoothing, averaging
the nearest 100 local values. In Figure 8 this average is plotted in red. With $dt$ positive the
exponent $\lambda_1(t)$ is uniformly positive, changing sign with the velocity reversal. The reversed
reference and satellite trajectories attract for a while, as expected, but only for a while, for a time of order
3 with $dt=0.01$ and 5 with $dt = 0.00078125$. Because integration errors are magnified exponentially, as
described by the Lyapunov exponent, $\lambda_1 \simeq 1.8$, there is a systematic timestep dependence of the time
``$t_{rev}(dt)$''. The exponent is reversed for a time proportional to the logarithm of the number of timesteps,
$\propto \ln(1/dt)$. Roughly speaking, $t_{rev}(dt)$ increases by about 0.4 for each halving of the timestep. See
Figure 9. A set of seven computations with $0.05 \leq dt \leq 0.0078125$ suggests the phenomenological relation
$$
e^{1.8t_{rev}(dt)} \simeq dt^{-?} \rightarrow 1.8t_{rev} \simeq - ?\ln(dt) \ .
$$
In Figure 9 $t_{\rm rev}$ varies from about 3 to about 5 as the timestep decreases by a factor of 64.

The logarithmic relationship between timestep and reversal time is consistent with the exponential
amplification of few-step integration errors. A similar rough relationship holds for the limit-cycle
problem of Figures 2-4. Capturing a reversed trajectory with visual accuracy up to a time $t_{rev}(dt)$
requires a simulation effort varying as $(t_{\rm rev}/dt)$.

We are pleased to offer a reward for further investigation of these problem types: a special \$1000
``Snook Prize'', in honor of our late colleague Ian Snook, who died in 2013. Application for this Prize
requires the submission of an appropriate acceptable electronic manuscript addressing shockwave
reversibility to {\tt CMST.eu} prior to year's end of 2023.

\begin{figure}
\includegraphics[width=3in,angle=-90.]{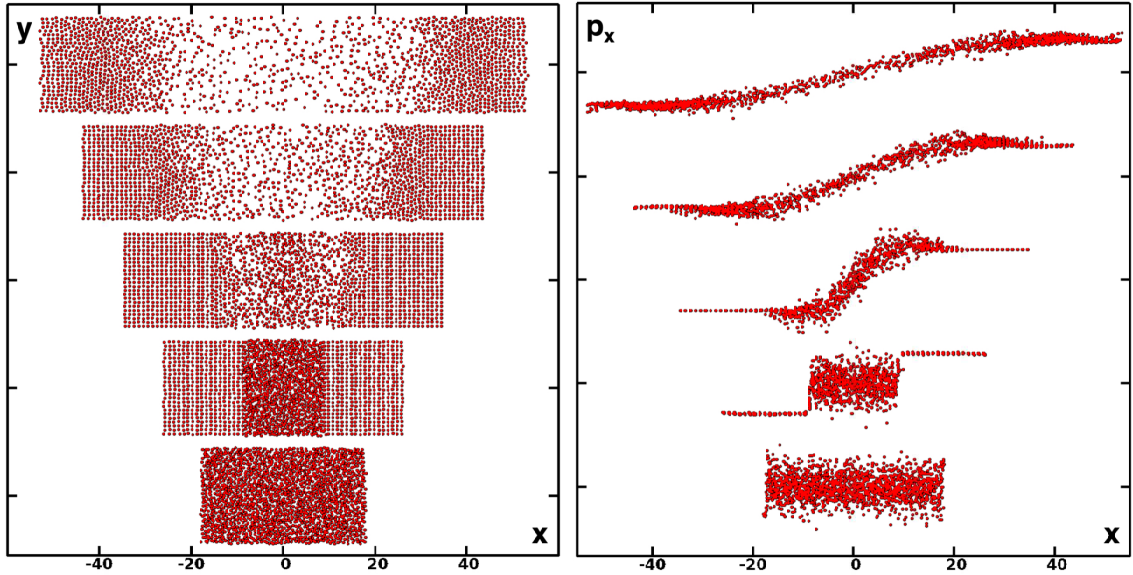}
\caption{
Velocity reversal following the twofold shock compression of 1680 soft
disks. The initial velocities $\pm 0.875$ at the left and right, resulted
in a headon collision. The initial particle coordinates at the base of this
Figure correspond precisely to the maximum-compression middle view of Figure 6.
The initial velocities were changed in sign so that the next-to-bottom
snapshot here corresponds closely to a reversal of the next-to-bottom view of
Figure 6. Soon after, Lyapunov instability, with $\lambda_1 \simeq 2$, prevents
additional reversed configurations at time 20, 30, and 40.  Just as in Figure 6
particle positions appear at the left with corresponding horizontal velocity
components at the right.
}
\end{figure}

\begin{figure}
\includegraphics[width=2in,angle=-90.]{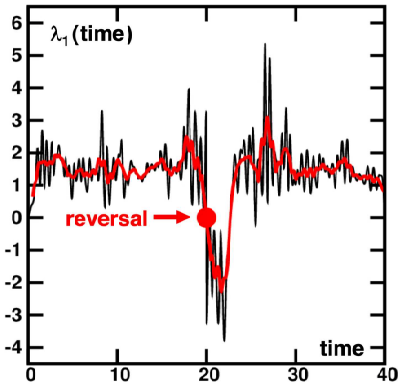}
\caption{
The local Lyapunov exponent for the time-reversal of Figure 6. The reversal, at time
20, is indicated by a filled red circle. The red curve corresponds to a smoothed
exponent averaged over unit time, 100 time steps with the step $dt = 0.01$.
Notice that the Lyapunov exponent is of order $\pm 2$ in the shocked dense
fluid or in the cold initial solid, suggesting an observable lack of reversibility
when the amplified roundoff error reaches the amplitude of Lyapunov instability,
a time of order 14 for $dt = 0.01$ :
$
0.01^5e^{\lambda_1(t)\times t}/120 \simeq 1 \rightarrow t \simeq 14 \ .
$
}
\end{figure}

\pagebreak

\begin{figure}
\includegraphics[width=2.4in,angle=-90.]{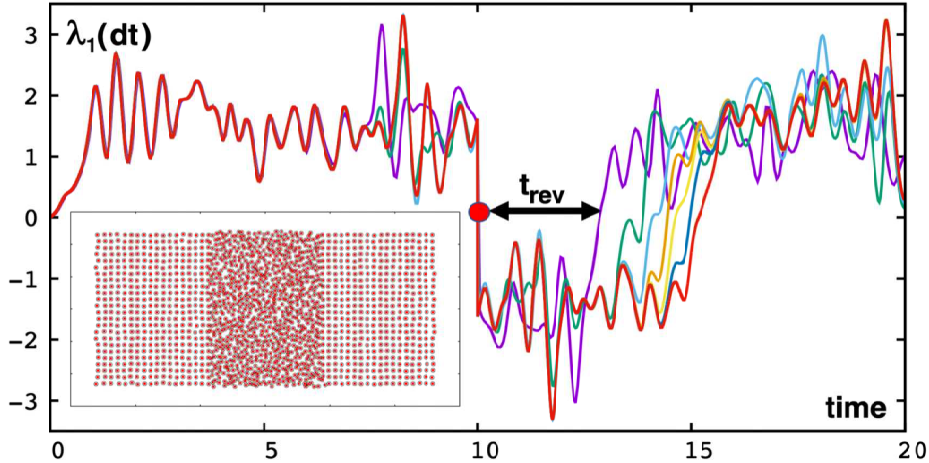}
\caption{
The largest Lyapunov exponent, $\lambda_1 \simeq 1.8$ for seven simulations with
$dt=0.1/2^n$ reflects the velocity sign change at time 10. The accuracy of the
reversal persists for a time $t_{\rm rev}$ varying as $\ln(1/dt)$, consistent 
with the Lyapunov amplification of integration errors to visible size. The inset
snapshot corresponds to the $t = 10$ configuration of 1680 soft-disk particles undergoing
twofold shock compression up until the reversal time indicated by the red dot.
}
\end{figure}

\begin{figure}
\includegraphics[width=3in,angle=-90.]{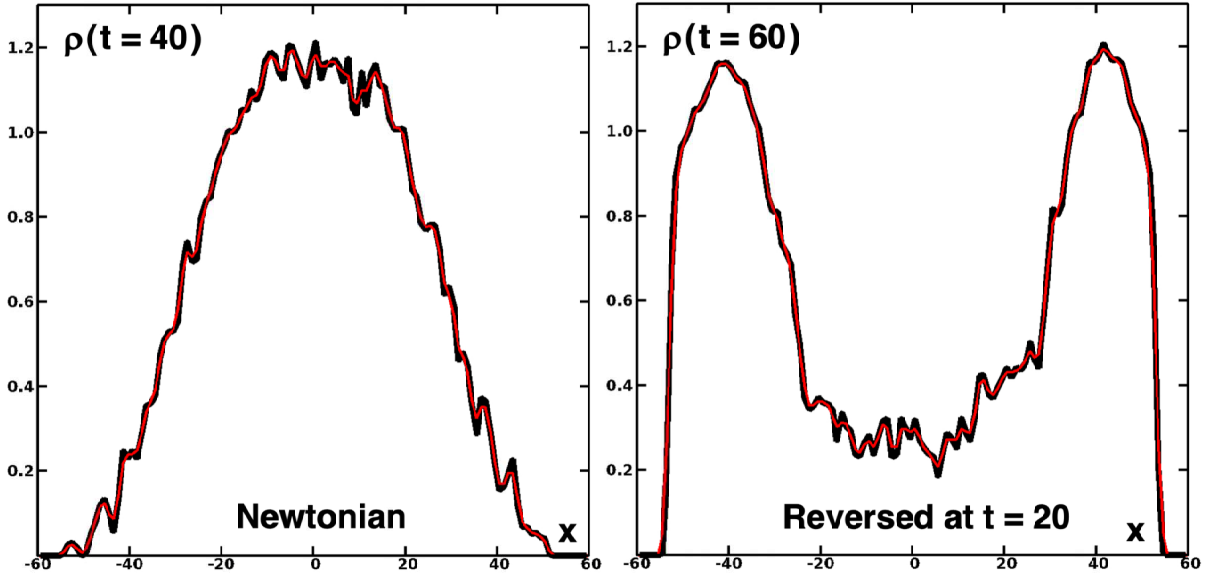}
\caption{
Smoothed-particle density profiles, Newtonian at the left and reversed at
the right. These profiles correspond to the topmost configurations of
Figures 6 and 7. Similar profiles of $P_{xx}$ and $P_{yy}$ show that the
pressure is nearly isotropic. The thick black and red lines correspond to
smoothed-particle ranges of 2 and 3 respectively.
}
\end{figure}

\section{Summary and Prognosis}
Our first problem, the thermostatted generation of a heat current, stabilized by
Nos\'e-Hoover mechanics, demonstrates that such dissipative examples, despite their
time reversibility, stabilize mass, momentum, and energy flows consistent with the
Second Law of Thermodynamics. The mechanism for this dissipative irreversibility
is the formation of fractal phase-space distributions with mirror-image fractal
pairs of distributions. With these pairs the attractor has probability one 
while the repellor has probability zero, even for a small system, the heat-conducting
oscillator. This same explanation of thermostatted irreversibility holds for
manybody systems, as was suggested in References 7 and 8.

The purely-Newtonian shockwave problem illustrates a different, but related, mechanism 
for irreversibility. Though ``mechanism'' is conceptual, its realization is necessarily
computational. Lyapunov instability destroys a time-reversed system's memory
by magnifying computational noise. Soon the amplified roundoff error becomes of order
unity. Our shockwaves simulations show that this time is of order $t \simeq$ 3 to 5, just a few
collision times. At that time the reversed flows are destabilized by the preponderance
of entropy-producing flows over entropy-reducing flows, giving a purely
Newtonian resolution of Loschmidt's paradox. No thermostatting control
variable is required.  But still it is likely that the probability of going
forward with an entropy-producing shock exceeds that of its reversal in a singular way. The
preponderance of states generating entropy, in the shock, over those which would
reduce it in the reversed shock, requires no modification of the Newtonian
motion equations.

In making the connection between computational simulations and continuum
mechanics smooth-particle averaging is an indispensable tool. To illustrate this
idea let us calculate density profiles $\rho(x)$ for the topmost snapshots
in Figures 6 and 7. Imagine the density of every particle to be spread out
in $x$ over a range $h$ according to a normalized weight function $w(r)$ with finite
range and two continuous derivatives:
$$
r<h \longrightarrow w(r,h) = [5/4hL_y][1 - (r/h)]^3[1 + 3(r/h)] \ .
$$
$$
[ \ {\rm Lucy's \ weight \ function} \ ]
$$
Summing the weights contributed by each of the particles to every grid point in
the set $\{x_g\}$ produces the smoothed density profiles shown in Figure 10 with
$h = 2$ and $h=3$. The grid spacing is unity so that every particle contributes
to four or 6 nearby grid points for these two values of $h$. We see that neither profile
matches the uniform density at the base of Figure 6. The regular square-lattice
structure near both ends of the hot shocked fluid (near $x=\pm 20$, at the base
of Figure 7 and the middle of Figure 6) is less susceptible to the smoothing
loss of memory due to Lyapunov instability.

Smooth-particle averaging can be applied to any of the atomistic functions
of coordinates and momentum. Plots of the pressure tensor indicate isotropy
with $P_{xx} \simeq P_{yy}$. Research into the details of the atomistic
distribution functions could elucidate further the mechanism responsible for
the exponentially greater density of phase flows obeying the Second Law to
those flouting it. There remains much to do in understanding the failure of
Loschmidt's cogent idea---questioning the ability of mechanics to provide an
understanding of thermodynamics.

\section{Acknowledgments}

We thank Professor Krzysztof Witold Wojciechowski (Pozna\~n) for his interest and support of
this work. He spotted typos in the first version of this manuscript, for which we
are grateful.  We also thank Professor Julien Clinton Sprott (Wisconsin) for stimulation during
our preparation of Reference 11.

\pagebreak

\end{document}